\documentclass{article}
\usepackage{spconf,amsmath,graphicx}
\usepackage{makecell}
\usepackage{amsthm,amsmath,amssymb}
\usepackage{mathrsfs}

\title{Adjacent Slice Feature Guided 2.5D Network for Pulmonary Nodule Segmentation}
\name{Xinwei~Xue$^{\star\dagger}$ \qquad Gaoyu~Wang$^{\star\dagger}$ \qquad Long~Ma$^{\star\dagger}$ \qquad Qi~Jia$^{\star\dagger}$ \qquad Yi~Wang$^{\star\dagger}$ \qquad}
\address{$^{\star}$DUT-RU International School of Information Science $\&$ Engineering, Dalian University of Technology\\
$^{\dagger}$Key Laboratory for Ubiquitous Network and Service Software of Liaoning Province}
\begin{document}
%
\maketitle
\begin{abstract}
More and more attention has been paid to the segmentation of pulmonary nodules. Among the current methods based on deep learning, 3D segmentation methods directly input 3D images, which takes up a lot of memory and brings huge computation. However, most of the 2D segmentation methods with less parameters and calculation have the problem of lacking spatial relations between slices, resulting in poor segmentation performance. In order to solve these problems, we propose an adjacent slice feature guided 2.5D network. In this paper, we design an adjacent slice feature fusion model to introduce information from adjacent slices. To further improve the model performance, we construct a multi-scale fusion module to capture more context information, in addition, we design an edge-constrained loss function to optimize the segmentation results in the edge region. Fully experiments show that our method performs better than other existing methods in pulmonary nodule segmentation task.
\end{abstract}

\section{Introduction}
In recent years, the segmentation of pulmonary nodule has become an increasingly important task for effectively helping doctors diagnose lung diseases. However, current pulmonary nodule segmentation methods are faced with a series of problems to be solved. 
 
The segmentation methods based on 3D network like V-Net~\cite{milletari2016v}, 3D U-Net~\cite{3D} always take the whole CT image as input, which brings lots of calculations because of too much redundant information. In addition, 3D segmentation methods may be affected by potential interlaminar motion artifacts and inconsistent resolution of different axes, which makes the advantages of 3D CNN is not obvious.

\begin{figure}[!htbp]
	\begin{center}
		\begin{tabular}{c@{\extracolsep{0.2em}}c}
			\includegraphics[width=.215\textwidth]{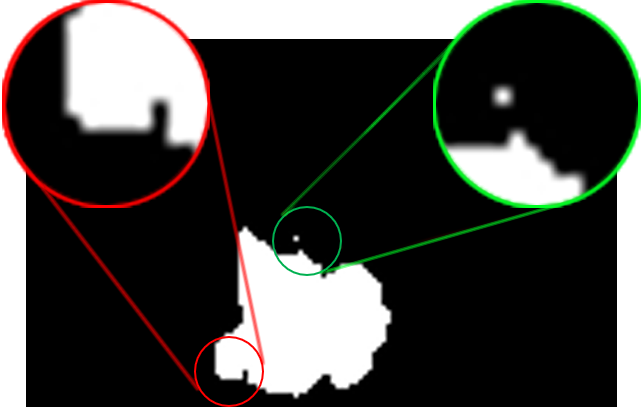}
			&\includegraphics[width=.215\textwidth]{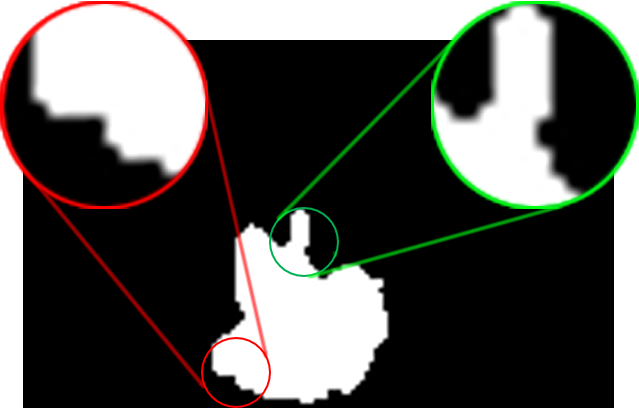}
			\vspace{-0.8mm}
			\\
			CE-Net\cite{gu2019net}&MVU-Net~\cite{wang2019automatic}
			\\
			\includegraphics[width=.215\textwidth]{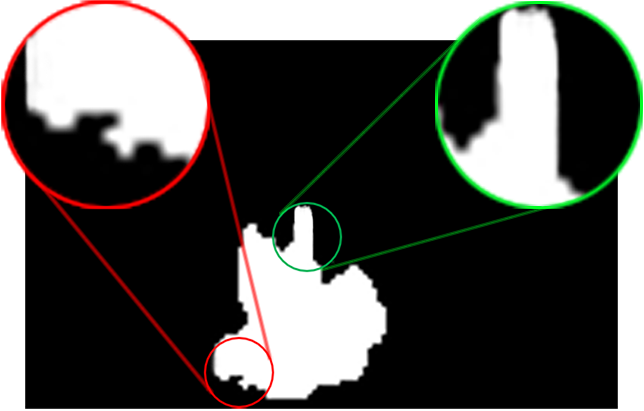}
			&\includegraphics[width=.215\textwidth]{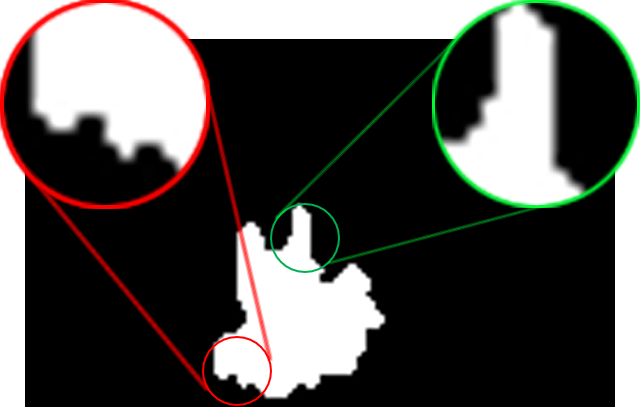}
			\vspace{-0.8mm}
			\\
			Ours&Ground Truth
			\\
		\end{tabular}
	\end{center}
	\vspace{-2em}
	\caption{Visual comparison on a challenging example among two state-of-the-art approaches and our method. Obviously, our method achieves the best visual expression with more delicate edge processing.
	\vspace{-1.8em}
	}\label{fig:1}		
\end{figure}
Generally speaking, 2D segmentation methods have less computation. The 3D CT image is cut evenly into 2D slices, and each slice is segmented by 2D model. Finally, the segmentation results of slices are combined into 3D result. However, most 2D medical segmentation methods, including U-Net~\cite{ronneberger2015u}, CE-Net~\cite{gu2019net}, ADAM-Net~\cite{wu2020automated} and PCA-U-Net~\cite{wang2021pca}, always ignore the spatial connection between slices, which leads to poor segmentation effect. 

The definition of 2.5D segmentation method is mainly divided into two types. One is the segmentation method using 2D convolution and 3D convolution at the same time such as MVU-Net~\cite{wang2019automatic}, and the other is the segmentation method fusing the features of adjacent slices on the basis of 2D segmentation of the target slice. 2.5D methods are intended to introduce more spatial information between slices without increasing the amount of data. However, most of the current 2.5D segmentation methods still cannot reach a satisfactory level in terms of experimental results. 
\begin{figure*}
	\begin{center}
		\centerline{\includegraphics[width=6.2in]{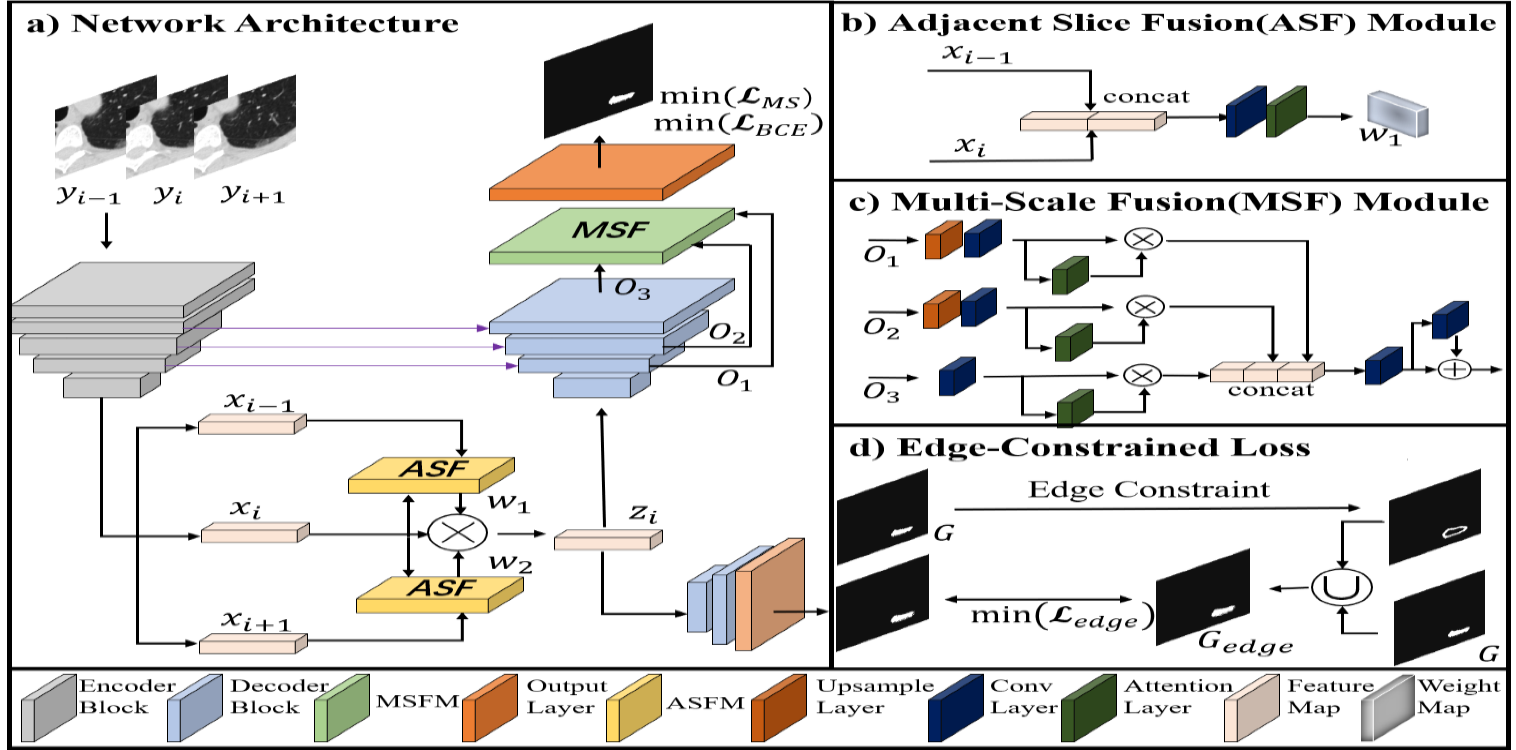}}
	\end{center}
	\vspace{-2.5em}
	\caption{The overall architecture of our proposed method. (a) Overall encoder-decoder network structure. (b) The structure of ASF module. (c) The structure of MSF module. (d) The diagram about how to construct ground truth with edge constraint. }
	\vspace{-1.5em}
\end{figure*}

To overcome the shortcomings of current medical segmentation methods for pulmonary nodule, we introduce the information of adjacent slices based on the 2D segmentation network, which is an effective way to improve spatial information. We improve the segmentation results by using more efficient multi-scale feature fusion module, and design new edge-constrained loss to optimize the processing of edge region. As shown in Fig. 1, our method performs better visual results with more accurate performance in the edge region. Finally, we conclude our contribution as the following folds:
\begin{itemize}
	\item We define an adjacent slice feature fusion model. This model clearly and concretely points out how to model the correlation of adjacent slices.
	\item We propose a medical segmentation network based on the adjacent slice feature fusion model. We design Multi-Scale Fusion(MSF) module to capture more context information and use edge-constrained loss function to refine the effect of segmentation in the edge region.
	\item We apply the proposed method in pulmonary nodule segmentation task.   Experiments show that the proposed method outperforms the state-of-the-art methods.
\end{itemize}

\section{Method}

\subsection{Explicit Modeling Across Slices}
Traditional 2D medical segmentation methods only focus on target slices, ignoring the information of adjacent slices and the spatial relationship between slices, resulting in unsatisfactory segmentation results. In order to introduce the information of adjacent slices, we design an adjacent slice feature fusion model. Our model can be formulated as
\begin{equation}
	z_{i}=\mathcal{A}(x_{i-1}, x_{i})\otimes x_{i}\otimes\mathcal{A}(x_{i}, x_{i+1}),
\end{equation}
where $\left\lbrace x_{i+k} = \mathcal{F}( y_{i+k})\right\rbrace_{k=-1,0,1}$ , $y_{i+k}$ means input images, $z_{i}$ is features map after fusion, $\mathcal{F}$ represents feature extraction module, $\mathcal{A}$ means Adjacent Slice Fusion(ASF) module with attention mechanism model. In this paper, we use resnet34~\cite{he2016deep} as the feature extraction model and CBAM~\cite{woo2018cbam} as the attention model.

Adjacent slices contain a lot of information related to the target slice, which can benefit the spatial connection between slices when using 2D convolution. Therefore, we choose to fuse adjacent slices and target slices at the feature level, but it is unreasonable to completely introduce the features of adjacent slices into the features of target slice. The main reason is that the information contained in adjacent slices is not all useful, the features of adjacent slices sometimes have a large gap with the target slice, which may mislead the segmentation process of target slice. In our model, we use attention mechanism to solve the above problem. As shown in Fig. 2, we first input adjacent slice $x_{i-1}$ and target slice $x_{i}$ and get their own feature maps through the feature extraction module, then we concatenate the feature map of adjacent slice and target slice on the channel axis, the new feature map after the concatenation can combine the information of two slices, and the difference with the target slice feature is not too large. Next, we will introduce the information of this feature map into the features of the target slice. The feature map is converted into a weight map after passing through the attention layer. This weight map $w$ contains the information of adjacent slices and target slices, which will be used as the information guide of the target slice, and the specific operation is that the weight map will be multiplied by the feature map of the target slice.

\subsection{Multi-Scale Fusion}
Multi-scale context fusion mechanism can improve the ability to obtain semantic information and geometric details by fusing different levels of receptive fields. Therefore, we design MSF module to make full use of the context information generated in the feature recovery module. The structure of MFS module is shown in Fig. 2, feature maps from different decoders first unify the size and number of channels through upsampling and convolution layers, then feature maps are concatenated together. We use attention mechanism~\cite{hu2018squeeze} to improve fusion efficiency in this process. Finally, the concatenated feature map restores the number of channels to the original feature map through convolution operation and input to the output layer. In addition, to boost the capture of non-linear features, we apply a series of convolution layers, BN and ReLU together with multiple residual connections for improved flow of information. 

\subsection{Edge-Constrained Loss Function}
Edge region segmentation is always a difficult problem in the field of image segmentation. In order to improve the segmentation effect of the edge region, We design the edge-constrained loss function, referring to SegMaR~\cite{jia2022segment}. First we need to define $G_{edge}$, $G_{edge}$ is computed via
\begin{equation}
	G_{edge}= G \cap A(\sigma,\theta,canny(G)),
\end{equation}
where $G$ represents binary mask(lable), $A(\bullet)$ is the Gaussian function with Gaussian blur $\sigma = 15$ and kernel size $\theta = 25$, $canny(\bullet)$ is the edge enhancement operation based on canny operator. As shown in Fig. 2, the feature map $z_{i}$ is input to a specific decoder branch, then the output image is compared with $G_{edge}$, $L_{edge}$ can optimize the encoder parameters through back propagation.
$L_{edge}$is defined as
\begin{equation}
	\begin{split}
		\mathcal{L}_{edge} =\lambda_{1}\mathcal{L}_{BCE}(O_{edge},G_{edge})
		+\lambda_{2}\mathcal{L}_{Dice}(O_{edge},G_{edge}),
	\end{split}
\end{equation}
where $O_{edge}$ represent the output from the decoder branch. $L_{BCE}$ means Binary CrossEntropy(BCE) loss function and $L_{Dice}$ means Dice coefficient loss function. $\lambda_{i}$ Indicates the corresponding parameter value.
\subsection{Multi-Scale Loss Function}
Multi-scale receptive field can help strengthen the global information and edge information, multi-scale loss function have been widely used in image processing methods. Refer to FPN~\cite{lin2017feature}, we compare the output of each decoder with the corresponding scale of ground truth. $L_{MS}$ is defined as
\begin{equation}
	\begin{split}
		\mathcal{L}_{MS} = \sum_{i=0}^2 \lambda_{i} \mathcal{L}_{BCE}(O_{i},G_{i}),
	\end{split}
\end{equation}
where $O_{i}$ represents the segmentation result of different scales output by different decoder and $G_{i}$ represents the mask of the corresponding scales. In addition, we also need a loss function $\mathcal{L}_{bin}$ to constrain the final output:
\begin{equation}
	\begin{split}
		\mathcal{L}_{bin} = \mathcal{L}_{BCE}(O,G)+\mathcal{L}_{Dice}(O,G),
	\end{split}
\end{equation}
 where $O$ represents final output. Our total loss function is
\begin{equation}
	\mathcal{L}_{total}=\mathcal{L}_{bin}+ \mathcal{L}_{edge}+\mathcal{L}_{MS}.
\end{equation}

\section{Experiments}
\begin{figure*}[!htbp]
	\begin{center}
		\begin{tabular}{c@{\extracolsep{0.2em}}c@{\extracolsep{0.2em}}c@{\extracolsep{0.2em}}c@{\extracolsep{0.2em}}c}
			\includegraphics[width=.185\textwidth]{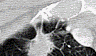}
			&\includegraphics[width=.185\textwidth]{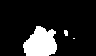}
			&\includegraphics[width=.185\textwidth]{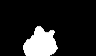}
			&\includegraphics[width=.185\textwidth]{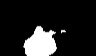}
			&\includegraphics[width=.185\textwidth]{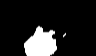}
			\vspace{-1mm}
			\\
			Input&U-Net(0.79)&CE-Net(0.82)&ADAM-Net(0.84)&CO-Net(0.84)
			\\
			\includegraphics[width=.185\textwidth]{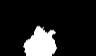}
			&\includegraphics[width=.185\textwidth]{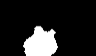}
			&\includegraphics[width=.185\textwidth]{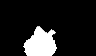}
			&\includegraphics[width=.185\textwidth]{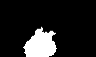}
			&\includegraphics[width=.185\textwidth]{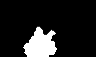}
			\vspace{-1mm}
			\\
			2.5D U-Net(0.88)&3D U-Net(0.86)&MVU-Net(0.91)&Ours($\mathbf{0.92}$)& Ground Truth
			\\
		\end{tabular}
	\end{center}
	\vspace{-1.8em}
	\caption{Qualitative results comparison on the LIDC-IDRI dataset with DSC score.
	\vspace{-2.5em}
	}\label{fig:3}		
\end{figure*}
\begin{figure*}[!htbp]
	\begin{center}
		\begin{tabular}{c@{\extracolsep{0.2em}}c@{\extracolsep{0.2em}}c@{\extracolsep{0.2em}}c@{\extracolsep{0.2em}}c@{\extracolsep{0.2em}}c@{\extracolsep{0.2em}}c@{\extracolsep{0.2em}}c}
			\includegraphics[width=.115\textwidth]{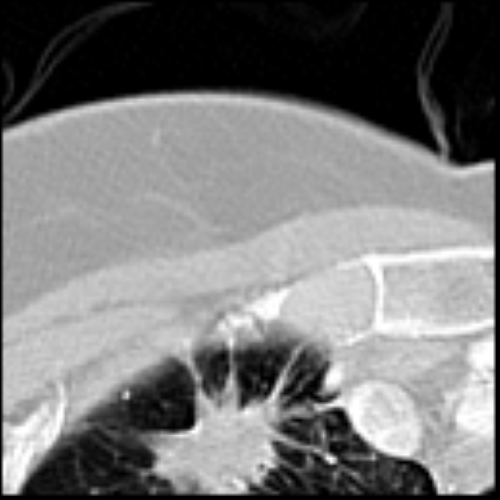}
			&\includegraphics[width=.115\textwidth]{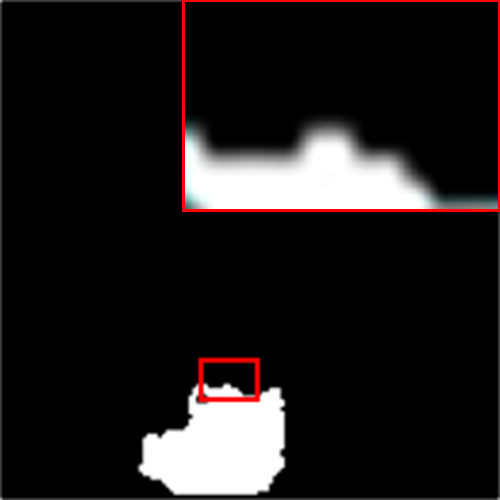}
			&\includegraphics[width=.115\textwidth]{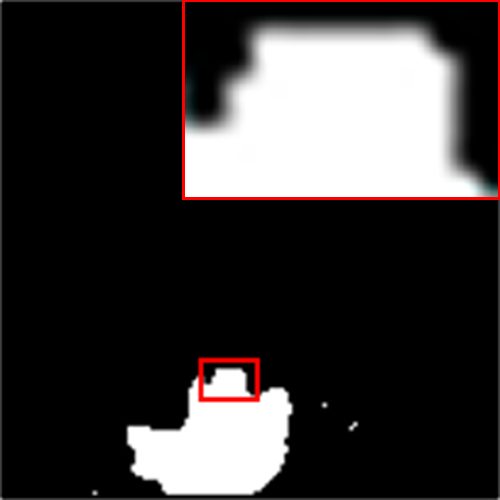}
			&\includegraphics[width=.115\textwidth]{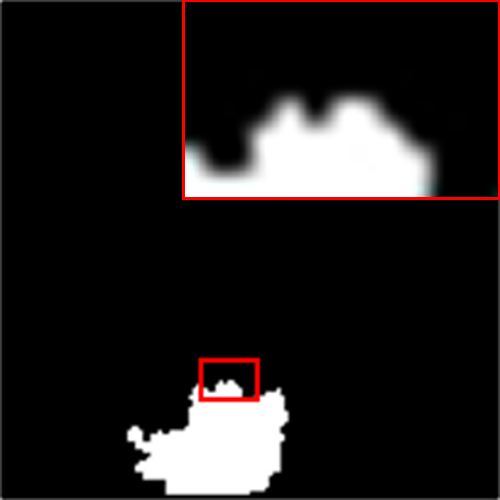}
			&\includegraphics[width=.115\textwidth]{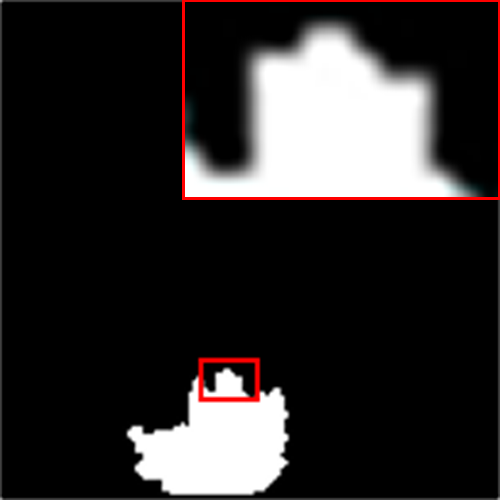}
			&\includegraphics[width=.115\textwidth]{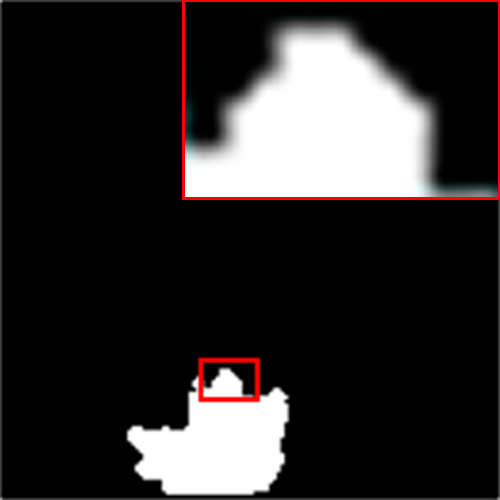}
			&\includegraphics[width=.115\textwidth]{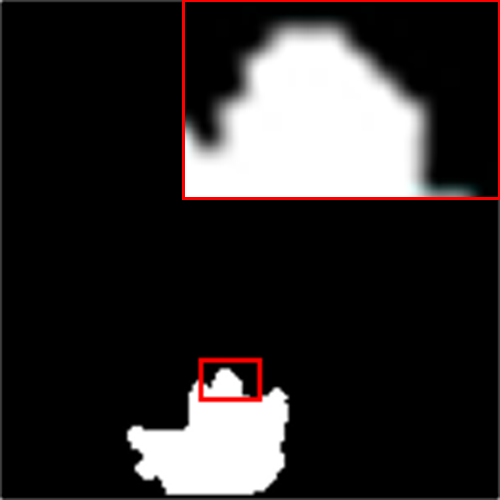}
			&\includegraphics[width=.115\textwidth]{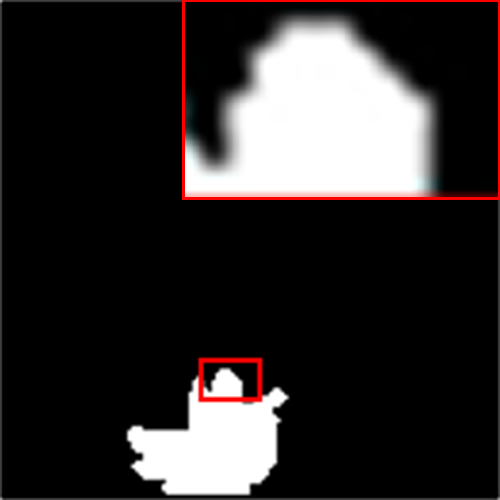}
			\\
			\includegraphics[width=.115\textwidth]{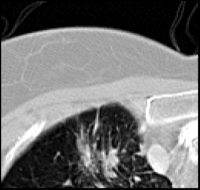}
			&\includegraphics[width=.115\textwidth]{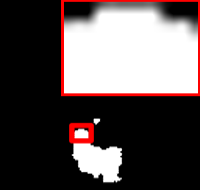}
			&\includegraphics[width=.115\textwidth]{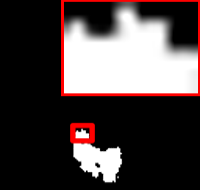}
			&\includegraphics[width=.115\textwidth]{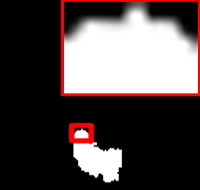}
			&\includegraphics[width=.115\textwidth]{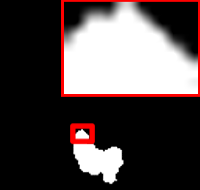}
			&\includegraphics[width=.115\textwidth]{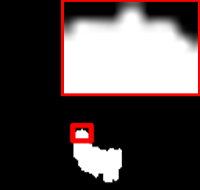}
			&\includegraphics[width=.115\textwidth]{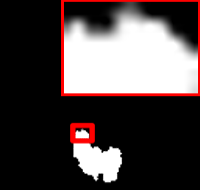}
			&\includegraphics[width=.115\textwidth]{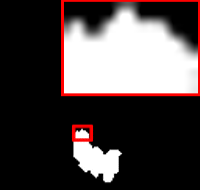}
			\vspace{-1mm}
			\\
			Input&U-Net&CE-Net&ADAM-Net&2.5D U-Net&MVU-Net&Ours& Ground Truth
			\\
		\end{tabular}
	\end{center}
	\vspace{-1.8em}
	\caption{More qualitative results comparison on the LIDC-IDRI dataset.
	\vspace{-2.5em}
	}\label{fig:4}		
\end{figure*}
We use LIDC-IDRI dataset, including 1018 CT images and corresponding nodule labels. First of all, we need to convert each CT image into 2D slice images, then process the slices refer to the method of Zhu et al.~\cite{zhu2018deeplung}. Generally speaking, the pulmonary nodules in each slice are always very small, therefore, we cut each slice into 16 sub images of equal size to improve the segmentation effect, these sub images will be used as the input of the network, and the corresponding output will be combined into a complete segmentation result, then the segmentation results of each slice will also be integrated into the final 3D segmentation result. In the experiments, we set 1000 groups of slices as training dataset and each group contains three adjacent slices.
\subsection{Comparison with Other Methods}
We compare the proposed method with some common methods applied to medical image segmentation, including U-Net~\cite{ronneberger2015u}, CE-Net~\cite{gu2019net}, ADAM-Net~\cite{wu2020automated}, CO-Net~\cite{liu2022co}, 2.5D U-Net~\cite{hu20182}, 3D U-Net~\cite{3D} and MVU-Net~\cite{wang2019automatic}. We use sensitivity (Sen), accuracy(Acc), IOU and DSC as indicators to evaluate the segmentation effect. To ensure the uniformity of the test results, after the final 3D segmentation results obtained by each method, we take the average score of the slices with nodes as the numerical result, referring to our training dataset. As shown in Table 1, experiments show that our method is better than other methods in quantitative results.

\begin{table}[h!]
	\vspace{-1.5em}
	\begin{center}
		\caption{ Quantitative results on the LIDC-IDRI dataset.}
		\begin{tabular}{ccccc} 
			\hline
			\textbf{Method} & \textbf{IOU} & \textbf{DSC} & \textbf{Sen} & \textbf{Acc}\\
			\hline
			2D U-Net & 0.705 & 0.827 & 0.898 & 0.896\\
			ADAM-Net & 0.748 & 0.876 & 0.931 & 0.974\\
			CO-Net & 0.788 & 0.881 & 0.936 & 0.978\\
			CE-Net & 0.742 & 0.869 & 0.929 & 0.966\\
			2.5D U-Net & 0.809 & 0.885 & 0.956 & 0.982\\
			MVU-Net & 0.818 & 0.895 & 0.961 & 0.989\\
			3D U-Net & 0.811 & 0.891 & 0.957 & 0.987\\
			Ours & $\mathbf{0.825}$ & $\mathbf{0.904}$ & $\mathbf{0.971}$ & $\mathbf{0.995}$\\
			\hline
		\end{tabular}
	\end{center}
\vspace{-1.8em}
\end{table}

Compared with other methods, it is obvious that our method has the highest score in each evaluation index, which means that our network can distinguish the background and foreground well, and the segmentation result is closer to the ground truth. The visual segmentation results are shown in Fig. 3 and Fig. 4, Fig. 3 shows more methods while Fig. 4 provides more comparison objects. From Fig. 4, obviously, our method has better segmentation effect in the edge region.
\vspace{-0.5em}
\subsection{Ablation Study}
We conduct the ablation studies to study the effectiveness of different components. Table 2 analyzes the impact of different modules and loss function on the segmentation network, Table 3 analyzes influence of different slice feature fusion model on segmentation results. $M_{0}$ in the table represents backbone, which is a 2D medical image segmentation network based on U-Net~\cite{ronneberger2015u}. $M1$-$M6$ respectively represents the methods of adding different component on the basis of backbone.
\begin{table}[h!]
	\vspace{-1em}
	\begin{center}
		\caption{Results of ablation study, analyzing the impact of different modules or loss functions on the segmentation result.}
		\begin{tabular}{cccccc} 
			\hline
			Method &ASF&  MSF &$\mathcal{L}_{edge}$&$\mathcal{L}_{MS}$& DSC \\
			\hline
			$M_{0}$ &  & & & & 0.871  \\
			$M_{1}$ & $\surd$ & & & &0.882  \\
			$M_{2}$ & $\surd$ &$\surd$ &  & &0.889  \\
			$M_{3}$ & $\surd$ &$\surd$ & $\surd$ & &0.892  \\
			$M_{4}$ & $\surd$ &$\surd$ & $\surd$&$\surd$ &0.904  \\
			\hline
		\end{tabular}
	\end{center}
\vspace{-1.5em}
\end{table}

According to Table 2, compared to the backbone, ASF module, MSF module, multi-scale loss function($\mathcal{L}_{MS}$), and $\mathcal{L}_{edge}$ have brought about 0.011, 0.007, 0012 and 0.003 numerical improvements respectively in terms of DSC evaluation indicators. The ablation experiment proves that the more components are selected, the better the segmentation performance can we achieve, which also means there is a positive impact between components.

We separately analyze the impact of the different slice feature fusion model in the ASF module. As shown in Table 3 and Fig. 5, by comparing numerical and visual results, we found that the fusion model we proposed is the best. 
\begin{table}[h!]
	\vspace{-1.5em}
	\begin{center}
		\caption{Ablation study about the impact of different slice feature fusion models. $F_{0}$ means only use target slices as input, $F_{3}$ represents the fusion model we proposed in the paper. $F_{1}$ represents model without attention layer, $F_{2}$ means that only the adjacent slices are used to get the weight map instead of first concatenating the target slice with the adjacent slices at the feature level.}
		\begin{tabular}{cccccc} 
			\hline
			Method & $F_{0}$ & $F_{1}$ &  $F_{2}$  & $F_{3}$ & DSC \\
			\hline
			$M_{0}$ & $\surd$ & & &&  0.871  \\
			$M_{5}$ &  &$\surd$ & & &0.864 \\
			$M_{6}$ &  & & $\surd$ & &0.876  \\
			$M_{1}$ &  & &  & $\surd$& 0.882  \\
			\hline
		\end{tabular}
	\end{center}
	\vspace{-1.5em}
\end{table}
\begin{figure}[!htbp]
	\begin{center}
		\begin{tabular}{c@{\extracolsep{0.3em}}c@{\extracolsep{0.3em}}c@{\extracolsep{0.3em}}c@{\extracolsep{0.3em}}c}
			\includegraphics[width=0.085\textwidth]{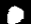}
			&\includegraphics[width=0.085\textwidth]{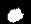}
			&\includegraphics[width=0.085\textwidth]{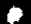}
			&\includegraphics[width=0.085\textwidth]{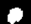}
			&\includegraphics[width=0.085\textwidth]{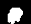}
			\vspace{-0.8mm}
			\\
			$M_{0}$&$M_{5}$&$M_{6}$&$M_{1}$& Ground Truth
			\\
		\end{tabular}
	\end{center}
	\vspace{-1.7em}
	\caption{ Influence of different slice feature fusion models on visual results, corresponding Table 3.
	\vspace{-1.8em}
	}\label{fig:img5}		
\end{figure}

\section{Conclusion}
This paper proposed an adjacent slice feature guided 2.5D medical segmentation network, the information of adjacent slices is introduced to enhance target slice features. We improved the segmentation effect by capturing context information and constraining edge regions, the abundant experiments fully demonstrated the superiority of our network in the pulmonary nodule segmentation task.

\bibliographystyle{IEEEbib}
\bibliography{refs}
\end{document}